\documentclass[pra,preprint,tightenlines,showpacs,twocolumn,10pt]{revtex4} 

\usepackage{epsfig} 
\usepackage{graphicx}
\usepackage{braket}
\usepackage{xcolor}
\usepackage[utf8]{inputenc} 
\usepackage{subfigure}

\usepackage{amsmath, amsfonts, amssymb, bm,mathtools}

\usepackage{soul,ulem,cancel}

\begin{document}  

\title{ Positronium breakup versus hydrogen ionization in collisions with fast charged projectiles: 
a comparative study }
 
\author{ B. Najjari$^1$, S. F. Zhang$^{1,3}$, X. Ma$^{1,3}$      
and A. B. Voitkiv$^2$ }
\affiliation{ $^1$ Institute of Modern Physics, Chinese Academy of Sciences, Lanzhou 730000, China \\ 
$^2$ Institute for Theoretical Physics I, Heinrich-Heine-University of D\"usseldorf, Universit\"atsstrasse 1, 40225 D\"usseldorf, Germany \\ 
$^3$ University of Chinese Academy of Sciences, Beijing 100049, China } 

\date{\today} 

\newcommand*\rfrac[2]{{}^{#1}\!\!/\!_{#2}}

\begin{abstract} 

We perform a comparative study of the breakup of positronium and ionization of atomic hydrogen 
by projectile-nuclei in the weak perturbation collision regime, $Z_p e_0^2/\hbar  \ll  v < c $ 
($v$ is the collision velocity, $Z_p$ the projectile atomic number, $e_0$ the elementary charge and $c$ the speed of light). In this regime the only principal difference between 
the collisions with these atomic systems 
lies in the masses of their positively charged constituents.  
We have shown that the corresponding mass effects strongly influence the spectra of the target fragments 
and the total cross sections. This influence manifests itself   
via i) the significantly smaller binding energy in positronium resulting in smaller momentum transfers necessary to break the system, ii) 
a strong constructive interference between the inelastic scattering of the projectile 
on the electron and the positron in collisions with positronium that also increases the chances for the breakup and iii) the "passive" role of the hydrogen nucleus caused by its heavy mass that prohibits hydrogen ionization to proceed via the interaction between the projectile and 
the nucleus. 
 
\end{abstract} 

\maketitle 

\section{ Introduction } 
 
The positronium (Ps) is a purely leptonic atom,  
which consists of an electron and its antiparticle, 
a positron, bound by an attractive Coulomb force.  
Ps is unstable because the electron and positron can annihilate.
The lifetime of Ps state-dependent. 
Annihilation in the ground state of Ps depends on its spin state:    
the singlet 1s ($^1$S$_0$) state (para-positronium)  
has a lifetime of $ \tau_{para} \approx 1.244 \times 10^{-10}$ s 
\cite{Ps-para-lifetime}, decaying predominantly into two gamma photons,  
whereas the triplet 1s ($^3$S$_1$) states (ortho-positronium) 
have a lifetime of $ \tau_{ortho} \approx 1.42 \times 10^{-7}$ s 
\cite{Ps-ortho-lifetime-theor}-\cite{Ps-ortho-lifetime-exp} 
(with a decay into three photons being  
the dominant annihilation channel). 

These lifetimes are several orders of magnitude larger 
than the typical orbiting time $\tau_{orb} \sim 10^{-16}$ s in Ps(1s). 
Therefore, in processes occurring on time scales much shorter 
than the lifetimes, the positronium may be practically regarded 
as a stable system which can participate in various interactions with matter, 
including biological media (see e.g. \cite{Ps-matter-1}-\cite{Ps-matter-2}), 
the interstellar gas (e.g. \cite{Ps-matter-3}-\cite{Ps-matter-4})  
and the Earth’s atmosphere \cite{Ps-matter-5}.   

The advent of positronium beams made 
collisions of Ps with with atoms in the range of Ps energies $\lesssim 400$ eV amenable to 
experimental investigation \cite{Ps-atom-review}.     
More recent advances in experimental techniques 
\cite{exper-with-Ps-1}-\cite{exper-with-Ps-3} 
have made it possible to produce cold positronium gases, 
paving the way to precise experiments with positronium. 

There is a number of experimental and theoretical studies of ionization and excitation 
occurring in collisions of Ps beams with atoms \cite{Ps-atom-review}, \cite{Ps-atom-1}--\cite{Ps-atom-8}. 
The different aspects of the breakup of Ps in collisions with photons 
(via Compton scattering and photo ionization)  
have been theoretically studied as well \cite{c-Ps-1}--\cite{c-Ps-5}. 
However, to our knowledge, the breakup of Ps by 
fast charged projectiles, which move with velocities much higher than 
the typical orbiting velocities in Ps, was studied neither experimentally nor theoretically.    

Therefore, in the present paper we shall theoretically explore 
the breakup of Ps by fast charged nuclei 
in the regime of weak perturbations, $Z_p e_0^2/\hbar v \ll 1$,  
where $Z_p e_0$ is the projectile charge ($Z_p$ is the atomic number, 
$e_0$ the elementary charge) and $v$ the collisions velocity,      
We shall also compare this process with ionization of the fundamental atomic system -- the hydrogen atom -- by the same projectiles.   
 
The hydrogen atom,   
due to its relative simplicity, represents an atomic 'test' system, for which 
calculations of the basic collision processes can be most easily performed 
and where the corresponding physical effects and features 
manifest themselves in the most simple way and, thus, 
can be easier understood. 

Like in the hydrogen atom, there are just two atomic particles in the positronium. 
However, compared to the former, the latter has 
two distinct features: i) the constituents of the positronium have equal masses;   
ii) the positronium is unstable because of annihilation. 

The point ii) is not relevant for the present study because 
the breakup of a positronium by fast charged projectiles 
in the regime of weak perturbations      
can efficiently proceed only provided the effective collision time 
does not exceed the typical transition time of the positronium. 
Since the latter is of the order the typical orbiting time in the positronium 
and, thus, is orders of magnitude shorter than the lifetimes of Ps,   
the annihilation has no direct influence on the breakup process.  

In contrast, the point i) is very important since the collision dynamics may significantly 
depend on the masses of the involved particles. 
In fast collisions, $v \gg e_0^2/\hbar $, 
the typical momentum transfers are orders of magnitude smaller than the momentum of the incident projectile-nucleus. As a result, the breakup/ionization cross sections do not depend on the projectile's mass. Moreover, in the regime of weak perturbations 
the projectile-target interaction can be treated within the first order of perturbation theory 
which means the practical absence of the charge sign effects (e.g., there will be essentially no difference between the spectra of electrons and positrons emitted in the breakup of Ps).  

Thus, the only principal difference between the breakup of Ps and ionization of hydrogen in such  collisions is that their positively charged constituents possess very different masses and the main goal of our study is to explore the impact of this difference -- of the target mass effects -- on the breakup/ionization processes.  

The paper is organized as follows. In the next section we obtain the transition amplitude and cross sections for the breakup of a two-body system by the impact of a high-energy charged projectile. 
In Section III we present our numerical results and their discussion.  
Section IV contains main conclusions. 

\section{ General consideration }

Let us consider an atomic system, which consists of two point-like particles labelled 
by indices $1$ and $2$.  We denote the charge and mass of particle $1$ by $ e_1 $ and $ m_1 $, respectively, and let ${\bm r}_1$ be coordinate of this particle with respect to 
the origin which we assume to rest in the laboratory frame \cite{origin}. 
Further, we denote 
by $ e_2 $ ($ e_1 \,\,  e_2 < 0$), $ m_2 $ and $\bm r_2$ the corresponding quantities 
for particle $2$. Let the two-body system interact with the projectile-ion with an atomic number 
$Z_p$ which is incident on the system with a velocity $\bm v =(0,0,v)$ along the 
$z$-axis. 

We assume that the initial velocity of the two-body system 
is much smaller than the projectile velocity $v$ and that the latter is high enough, 
$ v \gg Z_p v_0 $ where $v_0 = e_0^2/\hbar $, such that the collisions occur in 
the regime of weak perturbations. This also implies that 
the projectile velocity is much 
higher than the typical orbiting velocities ($ \sim v_0 $)  
of particles $1$ and $2$ in their initial bound state. The upper limit on 
the projectile velocity in our consideration is limited  
by the speed of light $c$. Thus, the collision velocity lies in the range 
$ Z_p e_0^2/\hbar \ll v < c$.     

We start our consideration of the ionization/breakup 
with the following remarks. 
Firstly, even in collisions of relativistic 
projectiles with such light targets, as hydrogen and positronium, 
the overwhelming majority of the target fragments 
has {\it nonrelativistic} energies 
\cite{ener-dist-e}, \cite{ener-dist-t} 
Therefore, we will employ  
a nonrelativistic description for the two-body target system.   
Secondly,  
since in the collisions under consideration the momentum transfers 
are negligibly small compared to the projectile momentum, 
the deflection angle of the projectile will be always extremely small. 
This enables us to begin the consideration 
with the semiclassical picture of the collision 
in which the projectile is described classically   
as moving in the laboratory frame along a straight-line trajectory  
$ \bm { \mathcal{ R } }(t) = \bm b +  \bm v t$,
where $ {\bm b} =(b_x; b_y; 0)$ is the transverse coordinate of the projectile 
(with respect to the origin), 
$ {\bm v} =(0; 0; v)$ the projectile velocity and
$t$ the time. When a semi-classical approximation is applied to collisions 
of ions with ordinary atoms both heavy particles (the ion and the atomic nucleus) are described classically (see e.g. \cite{E-M}, \cite{Eic}).    
However, for collisions with Ps such an approach is invalid due to the absence 
of a target nucleus and in our consideration the particles of the atomic two-body system 
(both Ps and H) will be treated quantum mechanically.   

The Schr{\"o}dinger equation for the two-body system 
interacting with the field of the projectile reads 
\begin{eqnarray}
i \frac{ \partial \Psi }{\partial t} = \hat{H} \Psi,  
\label{e1}
\end{eqnarray}
where semiclassical total Hamiltonian
$\hat{H} = \hat{H}(t)$ consists of 
the part $ \hat{H}_0 $, which describes the undistorted target,
and the interaction part, $ \hat{W}(t) $.

The Hamiltonian of the undistorted target is given by 
\begin{eqnarray}
\hat{H}_0 = \frac{ \hat{\bf p}_1^2  }{ 2 m_1 } +  %
\frac{ \hat{\bf p}_2^2  }{ 2 m_2 } + \frac{ e_1 e_2 }{ r_{12} },  
\label{e2}
\end{eqnarray}
where      
$ \hat{\bf p}_j = - i \hbar \bm \nabla_{\bm r_j} $ 
is the momentum operator for the $ j $-th atomic particle  
and $ {\bf r}_{12} = {\bf r}_1 - {\bf r}_2 $. 
 
The interaction of the target 
with the projectile reads 
\begin{eqnarray}
\hat{W} = \hat{W}_{I} + \hat{W}_{II} ,  
\label{int}
\end{eqnarray}
where 
\begin{eqnarray}
\hat{W}_I = \sum_{j=1}^2 \left( e_j \Phi_j %
- \frac{e_j}{2 m_j c} \left( \hat{\bm p}_j \cdot {\bm A}_j + %
{\bm A}_j \cdot \hat{\bm p}_j \right) \right),  
\label{int1}
\end{eqnarray}
and 
\begin{eqnarray}
\hat{W}_{II} = \sum_{j=1}^2 
\frac{ e_j^2 }{ 2 m_j c^2 } { \bm A}_j^2.   
\label{int2}
\end{eqnarray}
Here, $ \Phi_j $ and $ {\bf A}_j $ 
are the scalar and vector potentials, respectively, 
of the projectile field at the position of the $j$-th atomic particle. 
In the interaction (\ref{int}) we neglected the coupling of the spins 
of particles $1$ and $2$ to the projectile's magnetic field 
${\bm B} = \bm \nabla \times  {\bm A}$ 
since this part of the projectile-target interaction is comparatively 
very weak and may safely be omitted when the breakup/ionization of such light systems 
as Ps and hydrogen is considered. 
    
The field of the projectile can be described by 
the Lienard-Wiechert potentials which read 
(see e.g. \cite{E-M}, \cite{Jack}) 
\begin{eqnarray} 
\Phi_j & = & \frac{ \gamma \,\, Z_p e_0 }{\sqrt{ (\bm r_{\perp, j} -  \bm b )^2 + 
\gamma^2 (z_j - v t )^2 }} 
\nonumber \\ 
{\bm A}_j & = & \frac{ {\bm v} }{ c } \, \Phi_j,   
\label{L-W}  
\end{eqnarray} 
where $\gamma = 1/ \sqrt{1 - v^2/c^2 }$ is the Lorentz factor of the projectile, 
$Z_p$ is the projectile's atomic number 
and $e_0 > 0 $ is the elementary charge (the charge of the proton/positron).  

By introducing the center-of-mass $\bm R$ and 
relative coordinates $\bm r$ according to  
\begin{eqnarray} 
\bm R & = & \frac{ m_1 \, \bm r_1 \, + \, m_2 \, \bm r_2 }{ m_t }
\nonumber \\ 
\bm r & = & \bm r_1 - \bm r_2,  
\label{corrdinates-1}
\end{eqnarray}
where $m_t = m_1 + m_2 $ is the total mass of the two-body atomic system,  
the initial and final states are given by    
\begin{eqnarray} 
\psi_i(\bm R, \bm r, t) & = &  { e^{ i {\bm P_i \cdot \bm R/\hbar} } \over \sqrt{ V_a } } \, 
\varphi_i(\bm r) \, e^{ - i E_i t/\hbar },    
\label{in-state} 
\end{eqnarray} 
and 
\begin{eqnarray} 
\psi_f(\bm R, \bm r, t) & = &  { e^{ i {\bm P_f \cdot \bm R/\hbar} } \over \sqrt{ V_a } } \, 
\varphi_f(\bm r) \, e^{ - i E_f t/\hbar  },    
\label{fin-state} 
\end{eqnarray} 
Here,  $ \bm P_i $ and $E_i = \bm P_i^2/2 m_t + \varepsilon_i$ 
are the initial total momentum and energy, respectively, of the two-body system, 
and $\varphi_i(\bm r)$ is the initial internal state of this system 
with the corresponding energy $\varepsilon_i$ of the internal motion.
Further, $\bm P_f$, $ E_f = \bm P_f^2/2 m_t + \varepsilon_f $, $\varphi_f(\bm r)$ 
and $\varepsilon_f$ have meanings similar to the above quantities 
but refer to the final state of the two-body system. 

Within the first order of the perturbation theory 
in the projectile-target interaction 
the semiclassical transition amplitude is given by 
\begin{eqnarray} 
a_{fi}({\bm b}) = - \frac{ i }{ \hbar } \int_{- \infty}^{+ \infty} \!\!\! dt 
\int d^3 \bm R \int d^3 \bm r \,  
\psi_f^* \, \hat{W}_I \, \psi_i.   
\label{ampl-1}
\end{eqnarray} 
One should note that, 
since we treat the collisions wihtin the first order perturbation 
theory in the projectile-target interaction, the term $\hat{W}_{II}$ had to be omitted because  
its inclusion into the first order amplitude 
would be not self-consistent, resulting in 
an unphysical behaviour of the cross section at   
relativistic impact energies \cite{abv2007}. 

After long (but rather straightforward) calculations,  
for the amplitude (\ref{ampl-1}) we obtain  
\begin{eqnarray} 
a_{fi}({\bm b}) = - \frac{ i }{ \hbar v } \, \, \frac{ 8 \pi^2 \, Z_p e_0 }{ V_a \, {k'}^2 } 
\, \, e^{ - i \, \bm k_\perp \cdot \bm b } \, \delta(k_\parallel - \omega/v) \, \, (I - J).  
\label{ampl-2}
\end{eqnarray}
Here,  
\begin{eqnarray} 
I & = & I_1 + I_2 
\nonumber \\   
I_1 & = & e_1 \, \langle \varphi_f \vert e^{i \nu_1 \bm k \cdot \bm r} 
\vert \varphi_i \rangle  
\nonumber \\   
I_2 & = & e_2 \, \langle \varphi_f \vert e^{-i \nu_2 \bm k \cdot \bm r} \vert \varphi_i \rangle, 
\label{ampl-3}
\end{eqnarray}
and 
\begin{eqnarray}  
J & = & J_1 + J_2 
\nonumber \\ 
J_1 & = & e_1 \, \frac{ \bm v }{ c^2 } \cdot    
\Bigg\{ \!\! \frac{ \bm P_i \! + \! \bm P_f }{2 m_t} 
\langle \varphi_f \vert e^{i \nu_1 \bm k \cdot \bm r} \vert \varphi_i \rangle 
\nonumber \\ 
&& + \frac{ 1 }{ 2m_1 } \langle \varphi_f \vert \hat{\bm p}_{\bm r} 
\, e^{i \nu_1 \bm k \cdot \bm r} 
\! + \! e^{i \nu_1 \bm k \cdot \bm r} \hat{\bm p}_{\bm r} \vert \varphi_i \rangle 
\Bigg\} 
\nonumber \\ 
J_2 & = & e_2 \, \frac{ \bm v }{ c^2 } \cdot    
\Bigg\{ \!\! \frac{ \bm P_i \! + \! \bm P_f }{2 m_t} 
\langle \varphi_f \vert  e^{-i \nu_2 \bm k \cdot \bm r} \vert \varphi_i \rangle 
\nonumber \\ 
&& - \frac{ 1 }{ 2m_2 } \langle \varphi_f \vert \hat{\bm p}_{\bm r} \, e^{-i \nu_2 \bm k \cdot \bm r} 
\! + \! e^{-i \nu_2 \bm k \cdot \bm r} \hat{\bm p}_{\bm r} \vert \varphi_i \rangle \! \! \Bigg\},   
\label{ampl-4}
\end{eqnarray}
where $ \nu_j = \mu/m_j$ ($j = 1,2$), $\mu = m_1 m_2/m_t $ is the reduced mass, 
$\hbar \bm k = \bm P_f - \bm P_i$ 
and $ \hbar \omega = E_f - E_i$ are the momentum and energy, respectively, 
transferred to the two-body system in the collision 
(as viewed in the laboratory frame), 
$ \bm k = (\bm k_\perp, k_\parallel) $ with 
$ \bm k_\perp \cdot \bm v = 0$, $ k_\parallel = \bm k \cdot \bm v/v $, 
and $ \bm k' = (-\bm k_\perp, - k_\parallel/\gamma) $.  

The structure of the quantities $I$ and $J$ shows that the contributions to the transitions 
of the two-body system arising from the inelastic scattering of the projectile on particles $1$ and $2$ add up coherently (on the level of the amplitude). This means that the cross sections can be influenced by the interference effects between these two reaction pathways.   

As we have already mentioned, the "standard" semi-classical approximation, 
where not only an ion but also an atomic nucleus are described classically \cite{E-M}, \cite{Eic},  
cannot be applied for collisions with Ps where both particles are light, having the same mass. Therefore, our treatment of the collision combines 
a classical description of the projectile motion with a fully quantum description 
of the target. 

A consequence of such an 'intermediate' approach is that 
the amplitude (\ref{ampl-2}) has certain similarities 
with an amplitude obtained in a fully quantum treatment,   
where both the target and projectile are regarded as quantum objects. 
In particular, since we describe the center-of-mass motion of the target by plane waves 
(which spread over the whole interaction space with equal weight), 
the probability $ |a_{fi}(\bm b) |^2 $ does not depend  
on space \cite{imp-par}.  
A similar situation also arises in a fully quantum treatment, 
employing plane waves, 
and it is well known 
how to deal with it (see e.g. \cite{b-d}).  

We first absolutely square 
the amplitude (\ref{ampl-2}), rewriting 
the term $ \big( \delta( k_\parallel - \omega/v) \big)^2 $  
according to 
\begin{eqnarray} 
(\delta( k_\parallel - \omega/v)^2 & = & \delta( k_\parallel - \omega/v) \, 
\delta( k_\parallel - \omega/v)   
\nonumber \\ 
& = & \delta( k_\parallel - \omega/v)  \frac{1}{ 2 \pi } \, \,  
\lim_{L_\parallel \to \infty }  \int_{ - L_\parallel/2 }^{ + L_\parallel/2 } \! \! \! \! \! \! \! \! \! \! 
dz \, e^{ i (k_\parallel - \omega/v )  z } 
\nonumber \\   
& = & \lim_{L_\parallel \to \infty } \delta( k_\parallel - \omega/v) \frac{ L_\parallel }{ 2 \pi },  
\label{delta-fun} 
\end{eqnarray} 
where $L_\parallel$ is the longitudinal dimension of the interaction volume,    
obtaining       
\begin{eqnarray} 
\vert a_{fi} \vert^2 = \frac{ 32 \pi^3  }{ \hbar^2  v^2  } \, \, \frac{ Z_p^2  e_0^2 }{ V_a^2 \, {k'}^4 } \, \, \vert I - J \vert^2 
\,  L_\parallel  \, \delta(k_\parallel - \omega/v). 
\label{prob-1}
\end{eqnarray} 
The integration of $ \vert a_{fi} \vert^2$ over $d^2 \bm b = 2\pi \, b \, db$ from 
$b_{min} = 0 $ to $b_{max}$ yields 
\begin{eqnarray} 
\int \! \! d^2 \bm b \, \vert a_{fi} \vert^2 = \frac{ 32 \pi^3  }{ \hbar^2  v^2  } \, \, \frac{ Z_p^2  e_0^2 }{ V_a^2 \, {k'}^4 } \, \, \vert I \! - \! J \vert^2 
\,  V  \, \delta(k_\parallel \! - \! \omega/v),  
\label{prob-2}
\end{eqnarray} where $ V = \pi \, b_{max}^2 \, L_{\parallel} $ is the interaction volume.  
Dividing the right hand side of (\ref{prob-2}) by $n_a \, V$, where $n_a = 1/V_a  $ is 
the target density, we obtain the transition probability per one target atom. 
By taking into account the number of the final states of the target atom, 
$ \frac{ V_a d^3  \bm P_f }{ (2 \pi \hbar)^3 } \frac{ V_r d^3  \bm p }{ (2 \pi \hbar)^3 }$, where $\bm P_f$ is the final total momentum of the target, $\bm p$ and 
$V_r$ are the momentum and normalization volume, respectively, of the relative motion of the target fragments, we finally arrive at the cross section \cite{norm-vol}   
\begin{eqnarray} 
\frac{ d \sigma  }{ d^3 \bm P_f \, d^3 \bm p } & = & \frac{ 4 \, Z_p^2 e^2_0 }{ v^2 } \, 
\, \, \delta \Bigg( \! \! P_{f,z} - P_{i,z} - \frac{E_f - E_i}{v} \! \! \Bigg)  
\nonumber \\ 
&\times & \frac{ \vert I - J \vert^2}{\Bigg(\! (\bm P_{f,\perp} \! - \! \bm P_{i,\perp})^2 
+ \frac{(E_f \! - \! E_i)^2}{v^2 \gamma^2} \! \Bigg)^2}.  
\label{cr-1}
\end{eqnarray}   
We note that the cross section (\ref{cr-1}) is also obtained 
by using a fully quantum treatment, in which not only 
the center-of-mass motion of the target but also 
the projectile are described by plane waves 
and it is taken into account that the momentum transfers in the collisions 
are much smaller than the incident projectile momentum. 

The above cross section can also be rewritten as the cross section differential in 
the momenta $\bm p_1$ and $\bm p_2$ of particles $1$ and $2$, respectively, which reads  
\begin{eqnarray} 
\frac{ d \sigma  }{ d^3 \bm p_1 \, d^3 \bm p_2 } & = & \frac{ 4 \, Z_p^2 e^2_0 }{ v^2 } \, 
\, \, \delta \Bigg( \! \! p_{1,z} + p_{2,z} - P_{i,z} - \frac{E_f - E_i}{v} \! \! \Bigg)  
\nonumber \\ 
&\times & \frac{ \vert I - J \vert^2}{\Bigg(\! (\bm p_{1,\perp} + \bm p_{2,\perp} \! - \! \bm P_{i,\perp})^2 
+ \frac{(E_f \! - \! E_i)^2}{v^2 \gamma^2} \! \Bigg)^2},   
\label{cr-2}
\end{eqnarray}   
where $E_f = \frac{ p_1^2 }{ 2 m_1 }  + \frac{ p_2^2 }{ 2 m_2 } $ and 
the quantities $I$ and $J$, given by Eqs. (\ref{ampl-3})-(\ref{ampl-4}), 
have to be rewritten in terms of $\bm p_1$ and $\bm p_2$.  

It was already mentioned that the inelastic scattering of the projectile on particles $1$ and $2$ of the two-body system add up on the level of the amplitude which can result in interference effects in the cross sections. The inspection of (\ref{ampl-2})-(\ref{ampl-4}) shows that 
the interference effects are maximal 
when the particles $1$ and $2$ have equal masses, $m_1 = m_2$, and, in addition, the absolute values of their charges are the same, $|e_1| = |e_2|$. This is exactly the case for the breakup of Ps. 

The above two conditions also imply that in the weak perturbation regime of 
the collisions the spectra of the positronium fragments -- the electron and positron -- are identical.  For definiteness, when considering the breakup of Ps we 
shall talk about the electron spectra (which is also more convenient for a comparison with the electron emission from hydrogen). 

Unlike the positronium, the hydrogen atom consists of particles whose masses 
differ by more than three orders of magnitude. 
This has two consequences. First, 
the spectra of the hydrogen fragments -- the electron and nucleus -- will in general differ even in the weak perturbation regime.
Second, the interference effects in the ionization of hydrogen are negligibly small  
because the scattering of the projectile on the hydrogen nucleus practially does not lead to ionization.  
Indeed, as it follows from Eqs. (\ref{ampl-3})-(\ref{ampl-4}), the ionization via the interaction of the projectile with the hydrogen nucleus requires momentum transfers, which are roughly by three orders of magnitude  
larger than those necessary 
when the ionization is caused by the interaction of the projectile with the electron. However,  
the contribution of collisions with such huge momentum transfers to the atomic cross sections 
is negligible.

\section{ Numerical results and discussion } 

Before we proceed to discussing our numerical results, two remarks may be appropriate. 

First, within the regime of weak perturbations
the cross sections possess a simple dependence, $\sim Z_p^2$, on the projectile charge. Therefore, in what follows we shall present results for the cross sections, which are divided by $Z_p^2$. 

Second, there exist three hydrogen isotops, the protium $^1$H, the deuterium $^2$H (or D) 
and the tritium $^3$H (or T), which naturally occur in nature. In all of them the reduced mass $\mu = m_1 m_2/m_t $ practically coincides with the electron mass $m_e$ that makes  
the processes of their ionization by fast projectiles to be very similar. In particular, 
the spectra of electrons emitted in these processes are essentially identical. 
Moreover, even the angular and momentum spectra of the 
target nuclei will practically be the same. Keeping all this in mind,  
below, for the definiteness, we shall talk about the ionization of $^1$H (=H).    

\subsection{ Energy distributions of the taregt fragments } 

\begin{figure}[h!]
\vspace{-0.15cm}
\hspace{-0.5cm}  
\includegraphics[width=8.1cm]{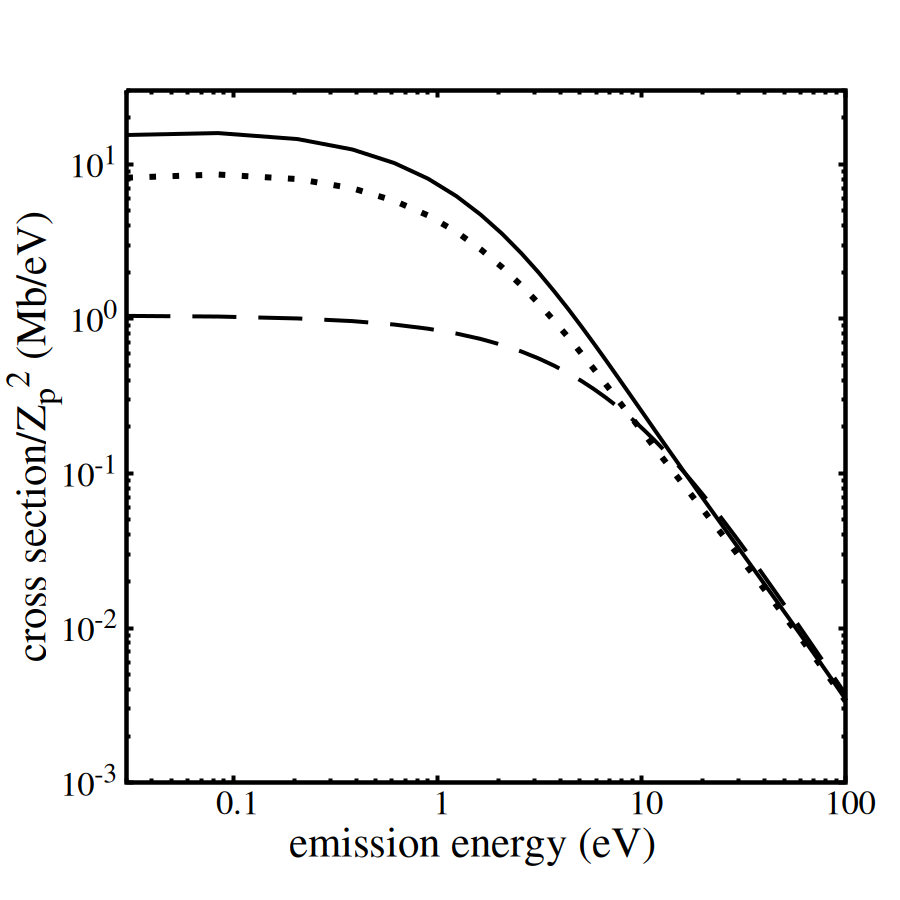} 
\vspace{-0.25cm} 
\caption{ The energy spectra of the emitted electrons in the breakup of Ps(1s) and 
ionization of H(1s) by $3$ MeV/u projectiles (the collision velocity $\approx 11 $ a.u.).  
Solid and dot curves: the emission in collisions with Ps(1s) calculated by including and neglecting, respectively, interference in the projectile scattering on the electron and positron. 
Dash curve: electron emission in collisions with H(1s).  }
\label{figure-energy-MeV}
\end{figure} 

\begin{figure}[h!]
\vspace{-0.15cm}
\hspace{-0.5cm}  
\includegraphics[width=8.1cm]{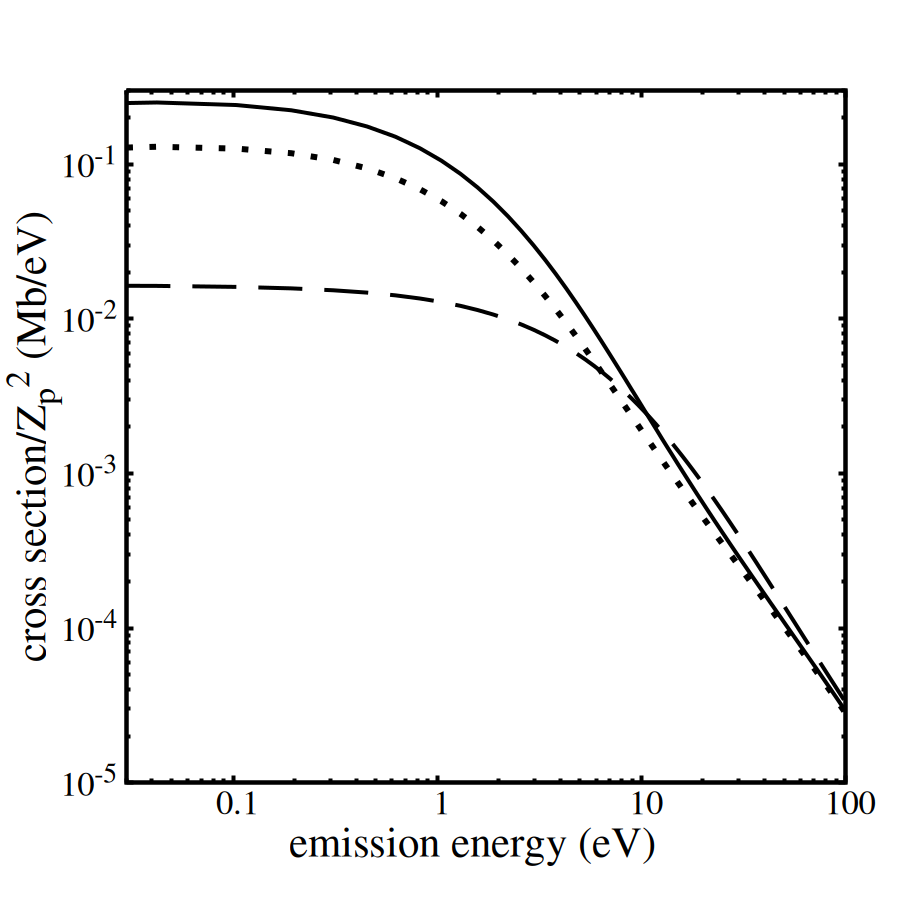} 
\vspace{-0.25cm} 
\caption{ Same as in figure \ref{figure-energy-MeV} but for $1$ GeV/u projectiles 
(the collision velocity $\approx 120$ a.u., the collisional Lorentz factor $\gamma \approx 2.08$).  }
\label{figure-energy-GeV}
\end{figure} 

In figures \ref{figure-energy-MeV} and \ref{figure-energy-GeV} 
we show the energy spectra of the emitted electrons  
in the breakup of Ps(1s) and ionization of H(1s) by $3$ MeV/u and $1$ GeV/u projectiles, respectively. 
Note also that these impact energies correspond to the collision velocity of 
$ v \approx 11 $ a.u. and $ v \approx 120$ a.u., respectively, and that at $1$ GeV/u  the corresponding collisional Lorentz factor $\gamma \approx 2.08$ already significantly exceeds $1$. 
   
Two main features are clearly seen in the figures. 
First, the intensity of the low-energy part of the spectrum in collisions with Ps 
greatly exceeds that in collisions with H.  
Second, by comparing the results for the breakup of Ps calculated by including and neglecting  the interference between the inelastic projectile scatterings on the electron and positron, we can conlcude that at low emission energies there is a strong constructive interference between these scattering channels. 

A comparison of the spectra at $3$ MeV/u and $1$ GeV/u also shows that 
at the higher impact energy the lower-energy part of the emission spectrum is significantly enhanced (in relative terms). This reflects an increasing dominance of the so called soft collisions -- collisions with (very) small momentum transfers -- in the processes of breakup/ionization by fast projectiles when the impact energy grows.    

A further inspection of the energy spectra in figures 
\ref{figure-energy-MeV} and \ref{figure-energy-GeV} leads to the following observations.  

First, in the lower-energy range (below few eV) the interference between the projectile scatteirng on the electron and positron increases the cross section for the breakup of Ps by nearly a factor of $2$. 
Taking the cross section, obtained by neglecting the interference, and
dividing it by $2$ would correspond to the contribution to the cross section from the projectile scattering just on one of the particles of Ps (either on the electron or on the positron). 
However, even such a reduced cross section still remains significantly larger than the cross section for the ionization of H. Since in the latter process
only the projectile scattering on the electron contributes (the proton is "passive" due to its very heavy mass, \cite{heavy-proton}),  it is plausible to attribute this remaining difference 
to the smaller binding energy in Ps.  
 
Second, the target reaction fragments with higher energy result 
from collisions with larger momentum transfers 
that diminishes the role of the interference effect in the breakup of Ps. 
Moreover, since larger emission energies are caused by the absorption 
of higher frequency components of the projectile's field,  
the larger binding energy in H now becomes favourable  
(this is similar to larger photoabsorption cross sections for
more tightly bound electrons, provided the photon energy is sufficient).
However, the effect of the larger binding energy in H is, to a large extent, 
compensated by the advantage of having two active particles in Ps. Therefore, 
the range of the large emission energies ($\gtrsim 10$ eV) contributes very 
little to the difference between
the total cross sections for the breakup of Ps and ionization of H.

\subsection{ Angular distributions of the target fragments } 

\begin{figure}[h!]
\vspace{-0.15cm}
\hspace{-0.5cm} 
\includegraphics[width=8.1cm]{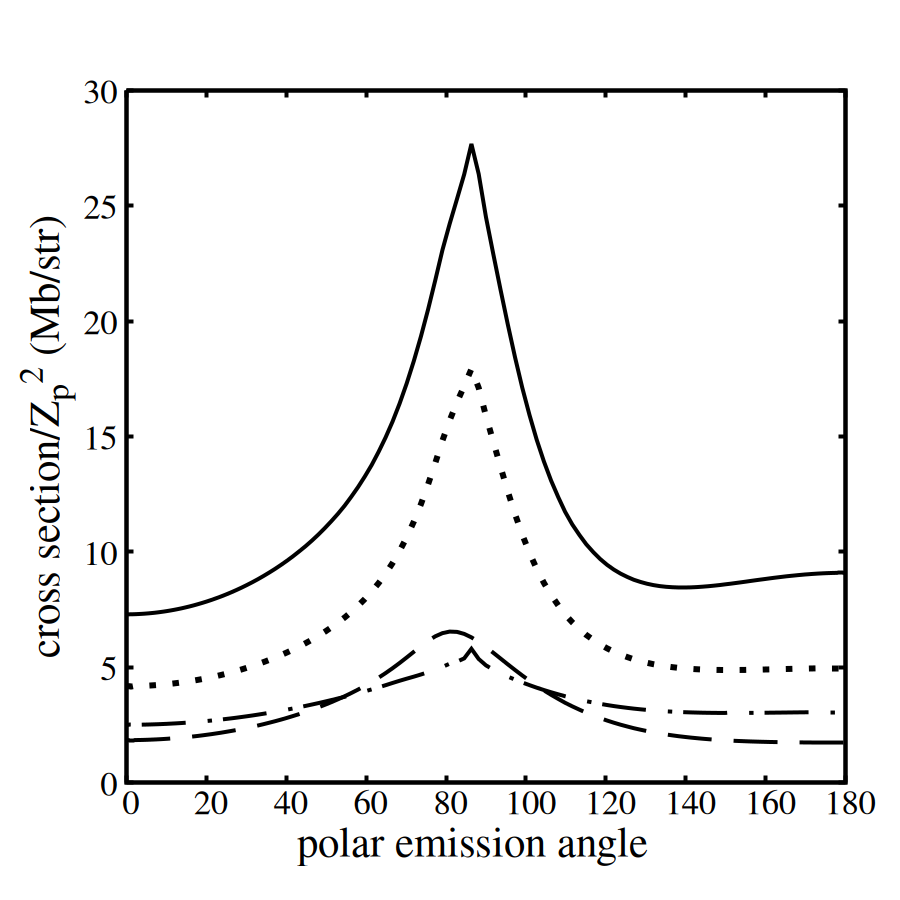} 
\vspace{-0.25cm} 
\caption{ The angular spectra of the target fragments in the breakup of Ps(1s) and 
ionization of H(1s) by $3$ MeV/u projectiles.  
Solid and dot curves: electron emission in collisions with Ps(1s) calculated by including and neglecting, respectively, interference in the projectile scattering on the electron and positron. 
Dash and dash-dot curves: electron and proton emission, respectively, in collisions with H(1s). }
\label{figure-angle-MeV}
\end{figure}

\begin{figure}[h!]
\vspace{-0.15cm}
\hspace{-0.5cm} 
\includegraphics[width=8.1cm]{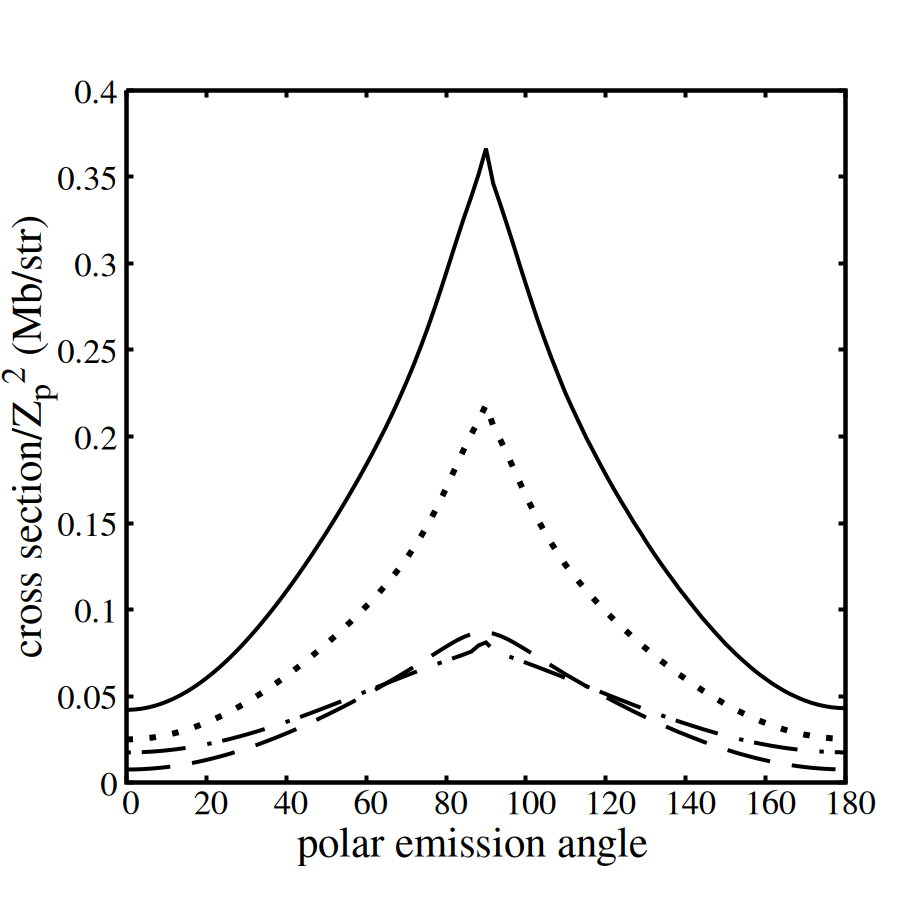} 
\vspace{-0.25cm} 
\caption{ Same as in figure \ref{figure-angle-MeV} but for an impact energy of $1$ GeV/u. }
\label{figure-angle-GeV}
\end{figure}

Figures \ref{figure-angle-MeV} and \ref{figure-angle-GeV} display the angular distributions of the target fragments at impact energies of $3$ MeV/u and $1$ GeV/u, respectively, 
represented by the cross section $\frac{d \sigma}{ \sin \theta d \theta}$,  where $\theta$ is the polar emission angle counted from the direction of the projectile motion $\bm v$.  

It is seen that in collisions at $3$ MeV/u the angular distribution of the emitted electrons  
possess a noticeable forward-backward asymmetry with a majority of them moving 
in the forward semi-sphere. 

The basic reason for this asymmetry is the positive value, $ \hbar k_\parallel > 0$, 
of the longitudinal component of the momentum transfer $ \hbar \bm k$ to the target in the collision. We also note that this asymmetry is more pronouced for the electron emission 
from hydrogen because i) the binding energy of hydrogen is larger than that of positronium and ii) it is the interaction between the projectile and the hydrogen electron which solely drives the process of ionization (and, thus, the momentum transfer to the target as a whole goes through the electron). At the same time, the angular distribution of protons is much less asymmetric since the proton participates only "indirectly" in the ionization process. 

In the collision with Ps the asymmetry in the angular distribution of the target fragments   
is less pronounced than in the spectrum of the electrons emitted from hydrogen 
because the values of $ \hbar k_\parallel $ are smaller for collisions with Ps due to the smaller binding energy. This asymmetry is however stronger than that in the spectrum of the protons since both particles in Ps "directly" participate in the breakup process. Also, in the breakup of Ps the constructive interference between the scattering on the electron and positron significantly enhances the angular spectrum of the target fragments for all emission angles. 

In collisions at $1$ GeV/u the asymmetry in the angular distributions becomes much weaker 
since the longitudinal component $ \hbar k_\parallel \sim 1/v $ of the momentum transfer 
is now much smaller. The breakup and ionization processes at such high impact energies are heavily dominated by collisions with very small momentum transfers. As a result, 
the interference effects in the breakup of Ps become somewhat stronger while in the ionization of hydrogen the difference 
between the angular distributions of the electrons and protons diminishes.    

To complete this subsection we note that the angular distributions 
of the emitted electrons in collisions with Ps and H qualitatively differ in the vicinity of 
their maxima: in the breakup of Ps this distrobution has a narrow cusp-like shape whereas 
in the ionization of H it is quite smooth. This difference can be traced back to the singularity 
of the Coulomb continuum function $ \sim 1/\sqrt{v_{12}} $ which arises when the relative velocity $v_{12}$
of the reaction fragments tends to zero. In the ionization of H, where the relative 
electron-proton velocity is practically equal to the electron velocity, in the angular distribution of the electrons the singularity in the wave function is compensated by 
the electron momentum $\sim p_e$ (which arises from the momentum space of the emitted electron: 
$ d^3 \bm p_e = m_e p_e d  \varepsilon_e  d \Omega_e$). Such a compensation, however, 
does not occur in the breakup of Ps because the electron (or positron) velocity 
now significantly differs from the relative electron-positron velocity.      

Moreover, the angular distribution of the protons in the ionization of H also possesses a similar cusp-like shape since the proton velocity significantly differs from 
the relative proton-electron velocity and, as a result, in the cross section  
the singularity in the continuum wave function is not compensated by the proton momentum. 

\subsection{ Longitudinal and transverse momentum distributions of the target fragments } 

\begin{figure}[h!]
\vspace{-0.15cm}
\hspace{-0.5cm} 
\includegraphics[width=8.1cm]{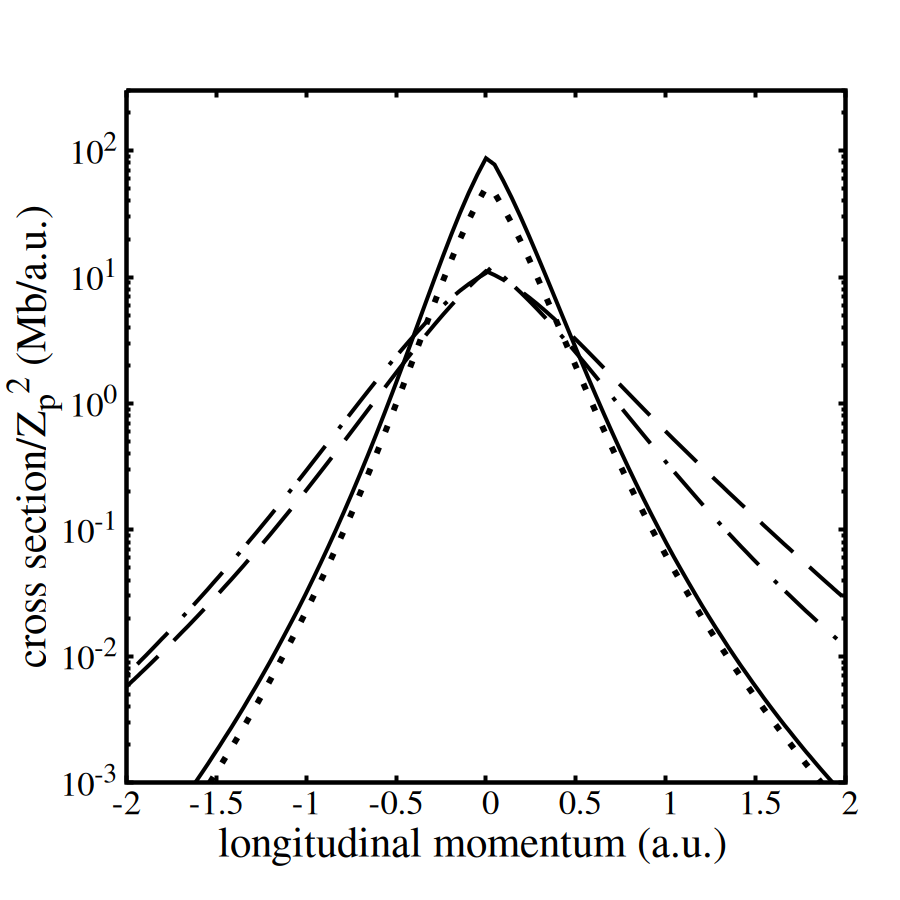} 
\vspace{-0.25cm} 
\caption{ The longitudinal momentum spectra of the target fragments in the breakup of Ps(1s) and 
ionization of H(1s) by $ 3 $ MeV/u projectiles.  
Solid and dot curves: emission in collisions with Ps(1s) calculated by including and neglecting, respectively, interference in the projectile scattering on the electron and positron. 
Dash and dash-dot curves: electron and proton emission, respectively, in collisions with H(1s). }
\label{figure-long-MeV}
\end{figure}

\begin{figure}[h!]
\vspace{-0.15cm}
\hspace{-0.5cm} 
\includegraphics[width=8.1cm]{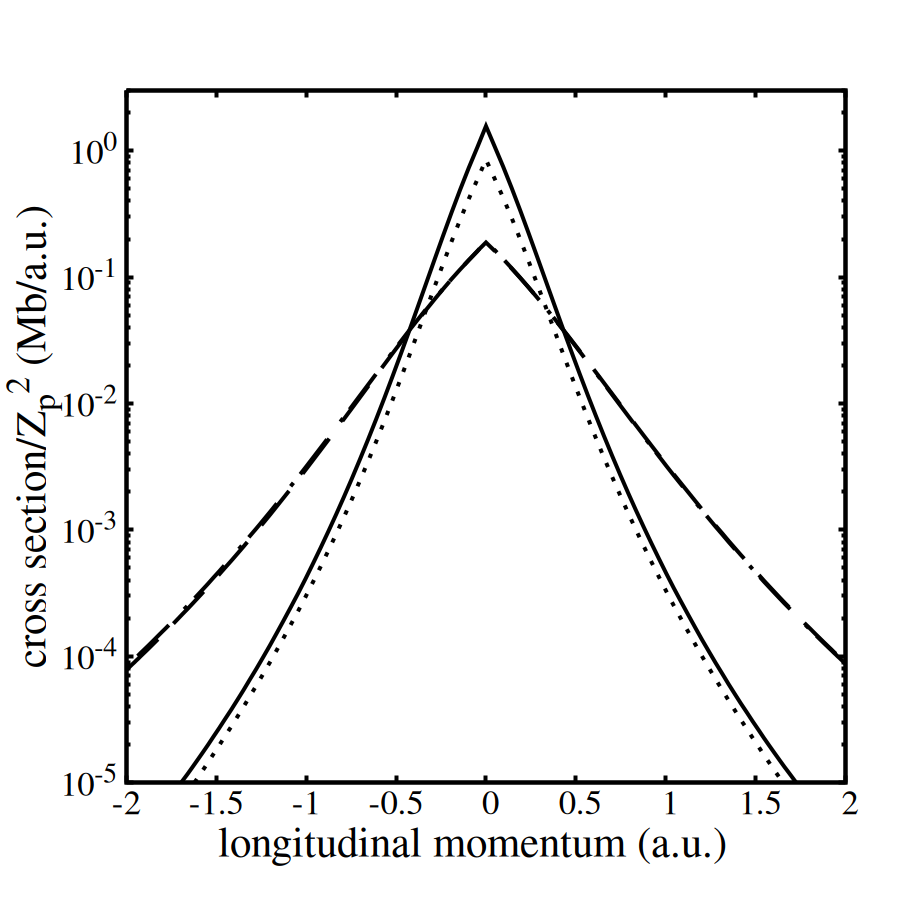} 
\vspace{-0.25cm} 
\caption{ Same as in figure \ref{figure-long-MeV} but for an impact energy of $1$ GeV/u. }
\label{figure-long-GeV}
\end{figure}

\begin{figure}[h!]
\vspace{-0.15cm}
\hspace{-0.5cm} 
\includegraphics[width=8.1cm]{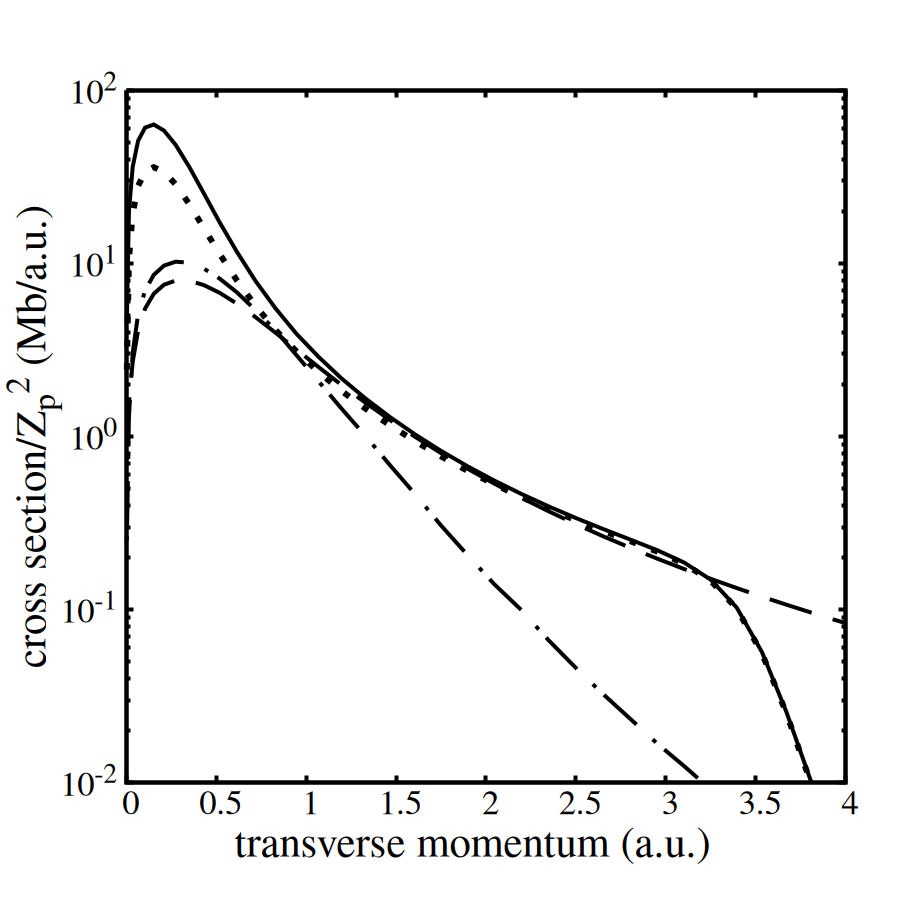} 
\vspace{-0.25cm} 
\caption{ The transverse momentum spectra of the target fragments in the breakup of Ps(1s) and 
ionization of H(1s) by $3$ MeV/u projectiles.  
Solid and dot curves: emission in collisions with Ps(1s) calculated by including and neglecting, respectively, interference in the projectile scattering on the electron and positron. 
Dash and dash-dot curves: electron and proton emission, respectively, in collisions with H(1s). }
\label{figure-trans-MeV}
\end{figure}

\begin{figure}[h!]
\vspace{-0.15cm}
\hspace{-0.5cm} 
\includegraphics[width=8.1cm]{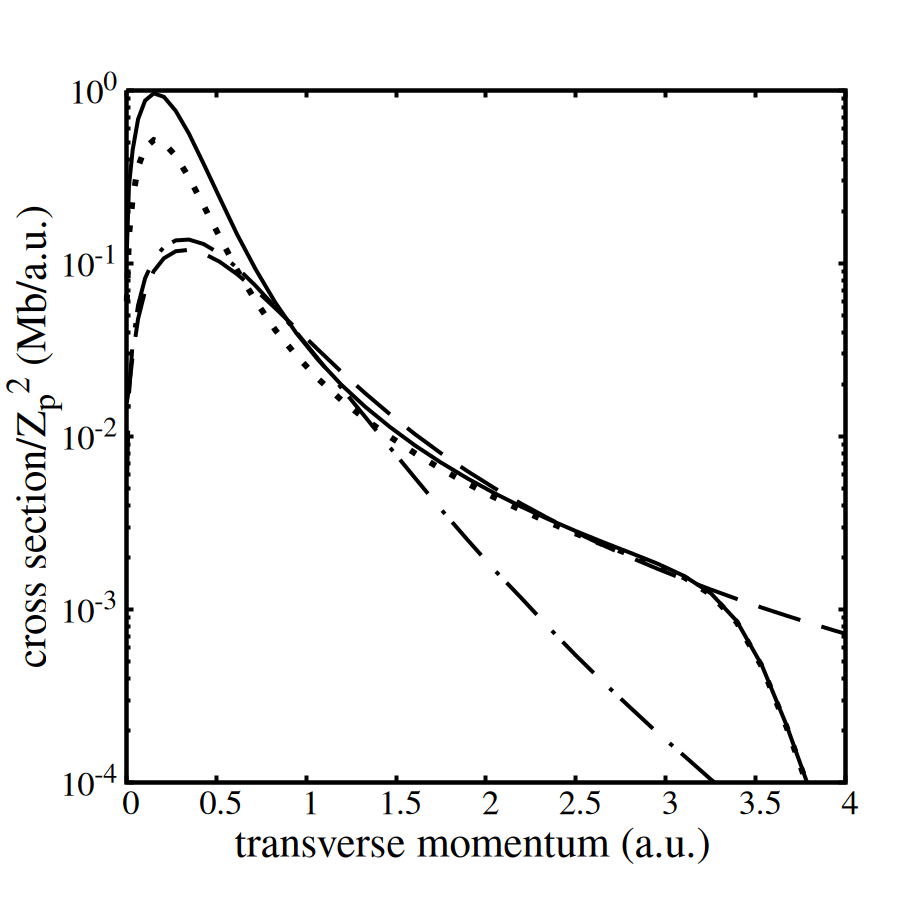} 
\vspace{-0.25cm} 
\caption{ Same as in figure \ref{figure-trans-MeV} but for an impact energy of $1$ GeV/u. }
\label{figure-trans-GeV}
\end{figure}

The exploration of the longitudinal, $ \frac{d \sigma}{ d  p_\parallel} $, 
and transverse, $ \frac{d \sigma}{ d  p_\perp} $, momentum distributions 
of the target fragments offers an additional perspective for looking into the collision dynamics. 
(Here, $ p_\parallel = \bm p \cdot \bm v/v$ and $ p_\perp = \sqrt{p^2 - p_\parallel^2}  $ are 
the longitudinal and transverse components, respectively, of the momentum $\bm p$ of the fragment.)
These distributions are presented in figures \ref{figure-long-MeV} - \ref{figure-trans-GeV} for 
impact energies of $ 3 $ MeV/u and $ 1 $ GeV/u. 

The longitudinal momentum distributions basically "confirm" and further stress 
the features observed in the angular distributions. 
Namely, in collisions at the smaller impact energy 
($ 3 $ MeV/u) i) there is a rather strong forward-backward asymmetry in the electron emission in collisions with H(1s), ii) this asymmetry is much weaker for the protons, which reflects the fact that the interaction between the projectile and the proton does not lead to ionization, and iii) this asymmetry is also significantly weaker for the target fragments in the breakup of Ps(1s) that is related to the smaller binding energy of this system resulting  in smaller values of the longitudinal momentum transfers in collisions with Ps. We also note that the broader longitudinal distributions of the target fragments in 
the process of hydrogen ionization are related to a broader Compton profile of the hydrogen atom. 

In collisions at much higher energy ($1$ GeV/u) the longitudinal momentum transfers in the collisions, which are proportional to $1/v$, are already much smaller than at $3$ MeV/u. As a result, 
the longitudinal momentum spectra become almost symmetric.   

We also note that all the longitudinal momentum distributions possess a sharp (cusp-like) maximum at $ p_{\parallel} = 0 $. This is a well known feature of atomic ionization in fast collisions (see e.g. \cite{ener-dist-t}). Its origin  
is the singularity of the Coulomb wave function for the relative motion of the target fragments at zero relative velocity. Unlike in the angular distributions, now this singularity is not compensated even for the electron emission from a heavy atom.

\vspace{0.35cm} 

The transverse momentum distributions of the target fragments for the same impact energies of the projectile are shown in figures \ref{figure-trans-MeV} - \ref{figure-trans-GeV}. The inspection of these figures shows that the main features in the breakup of Ps and ionization of H, which were discussed above, are also clearly seen in the transverse momentum distributions. In particular, we observe a strong constructive interference 
in the breakup of Ps at small momenta of its fragments and also that, with the momentum increase, 
this intereference eventually vanishes. Further, a more passive role played by the proton in 
the ionization of hydrogen, which leads to a significantly narrower momentum spectrum of this particle compared to that of the electron, is also clearly seen in these figures. 
We also observe that the breakup of Ps 
leads to a much stronger peak at low momenta of the positronium fragments 
compared to that of the electron in the ionizaton of H.  
Finally, figures \ref{figure-trans-MeV} - \ref{figure-trans-GeV} demonstrate a 
relative enhancement of the lower-momentum (lower-energy) part of the emission in collisions at the higher impact energy. 

\subsection{ The total breakup/ionization  cross sections } 

The total cross sections for the breakup of Ps and ionization of H can be obtained by 
performing the (multi-dimensional) numerical integration of the cross sections 
(\ref{cr-1}) - (\ref{cr-2}). 

It is known, however, that for single ionization 
occurring in high-energy collisions, one can derive simple 
(yet accurate) formulas  
for the total ionization cross section.      

The total cross section for hydrogen ionization 
from the ground state was derived long ago by H. Bethe \cite{Bethe} using 
the first order of perturbation theory in the projectile-target interaction    
(the so called Bethe-Born approximation, \cite{Bethe}, \cite{Inokuti}). 
This cross section, being written in atomic units ($\hbar = m_e = e_0 = 1$),  
reads 
\begin{eqnarray} 
\sigma_{\rm H} = 7.11 \, 
\frac{ Z_p^2 }{ v^2 } \,  
\Bigg( \ln(9.1 \,  v ) + \ln \gamma  - \frac{v^2 }{2 c^2 } \Bigg) 
\label{H-total}. 
\end{eqnarray}

For collisions with positronium 
a similar consideration (performed in the present study) 
yields for the total cross section for the breakdown from the ground state  
\begin{eqnarray} 
\sigma_{\rm Ps} = 28.45 \, 
\frac{ Z_p^2 }{ v^2 } \,  
\Bigg( \ln(5.6 \,  v ) + \ln \gamma  - \frac{v^2 }{2 c^2 } \Bigg) 
\label{Ps-total}. 
\end{eqnarray}
(Formula (\ref{Ps-total}) is also written in atomic units.) 

\begin{figure}[h!]
\vspace{-0.15cm}
\hspace{-0.5cm}  
\includegraphics[width=8.1cm]{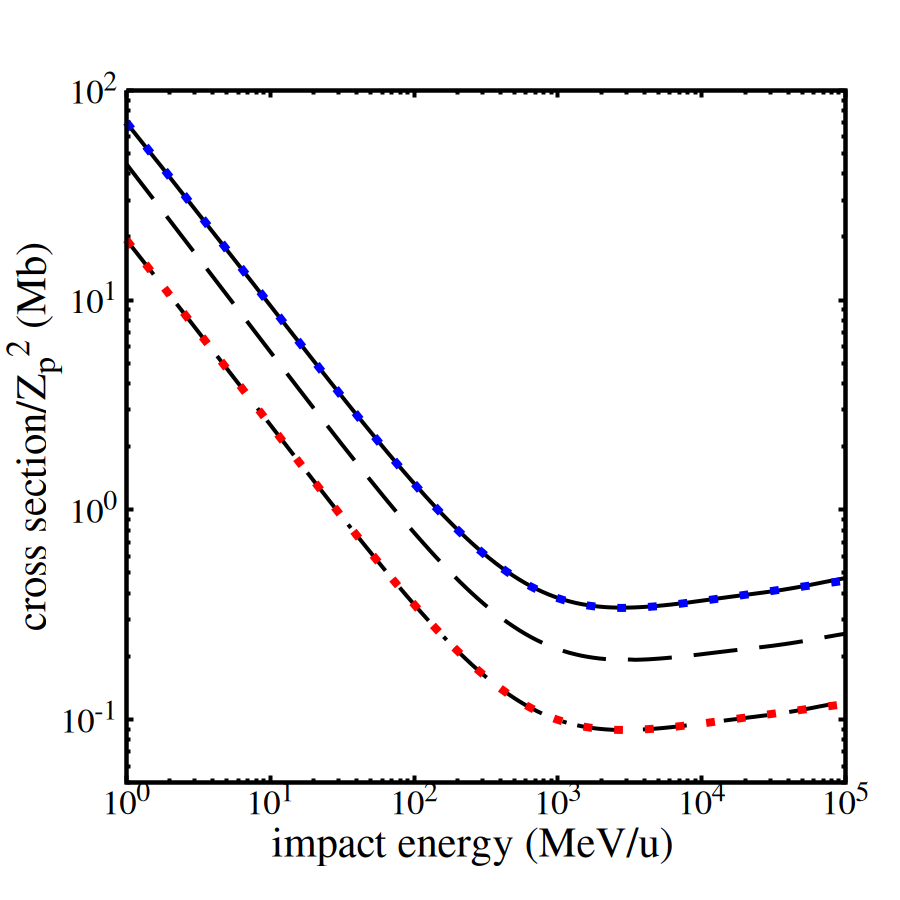} 
\vspace{-0.25cm} 
\caption{ The total cross sections for the breakup of Ps(1s) and 
ionization of H(1s) as a function of the impact energy. 
Solid and dash curves: the breakup of Ps(1s) calculated by including and neglecting, respectively, the interference effect in the inelastic scattering of the projectile on the electron and positron. 
Dash-dot curve: the ionization of H(1s). Blue and red points are calculations for 
ionization of H(1s) and breakup of Ps(1s) 
by using Eqs. (\ref{H-total}) and (\ref{Ps-total}), respectively. }
\label{figure-total}
\end{figure} 

Figure \ref{figure-total} displays the results for the total cross sections 
for the impact ionization of H(1s) and the breakup of Ps(1s).  
They were obtained i) by the numerical integration of the cross section 
(\ref{cr-1}) (or (\ref{cr-2})) and ii) by using expressions 
(\ref{H-total}) and (\ref{Ps-total}). It follows from these results 
that the cross section for Ps breakup is about $4$ times higher than the cross section for H ionization.  It is also seen in the figures that both formulas  
(\ref{H-total}) and (\ref{Ps-total}) 
yield results which are in excellent agreement with the numerical computation 
of the total cross sections.  

An impact energy of $1$ GeV/u ($v/c \approx 0.88$, $\gamma \approx 2.08$) 
(which is one of the two collision energies considered in the previous subsections) 
formally belongs already to the relativistic domain of the impact energies. 
However, the relativistic effects in the breakup/ionization of so weakly bound systems, 
like hydrogen and positronium, remain very modest even at this impact energy 
\cite{footnote}.

The latter can be easily seen in the structure of 
the cross sections (\ref{H-total}) and (\ref{Ps-total}), where the relativistic effects 
are described by the second and third terms in the parenthesis. It follows from these expressions that in order for these effects to significantly influence the cross section, the collisional Lorentz factor $ \gamma $ has to be not much less than the collision velocity 
(measured in atomic units). 
This is illustrated in figure \ref{figure-total} where the significant increase in the cross section, caused by the relativistic flattening of the elecrromagnetic field of the projectile, 
is observed only at impact energies above $10$ GeV/u. 

\section{ Conclusions }

We have presented results of the comparative study of the breakup of positronium and ionization of atomic hydrogen by fast projectile-nuclei in the regime of weak perturbations,    
$ Z_p e_0^2/\hbar  \ll \ v < c$, where the projectile-target interaction 
can be treated within the first order of perturbation theory and where  
the physics of atomic collisions becomes relatively simple.  
Within this collision regime the cross sections depend on the projectile charge as $(Z_p e_0)^2$, the (sign) charge effects are absent which, in particular, 
means that  
the spectra of electrons and positrons in the breakup of Ps become identical.  
Also, the cross sections do not depend on the mass of the projectile. 

Then the only principal difference between 
the collisions with these two fundamental two-body atomic systems 
lies in very different masses of their positively charged constituents. 
Consequently, the main focus of our study was on the influence 
of the corresponding mass effects on the breakup/ionization processes. 

We have shown that the mass effects manifest themselves via 
i) an overally strong constructive interference between the inelastic scattering of the projectile 
on the electron and the positron that increases the chances for the breakup of Ps, 
ii) the significantly smaller binding energy of Ps that results in smaller momentum transfers necessary to breakup the system and thus also increases the cross sections,  
and iii) the "passive" role of the hydrogen nucleus in the collisions 
caused by its very heavy mass that prohibits  
the ionization of hydrogen through the interaction between the projectile and the nucleus.     

We have explored in some detail the energy, angular and momentum spectra of the target fragments as well as the total cross sections, showing in particular that  
the spectra of electrons  
emitted in collisions with hydrogen and positronium very significantly 
differ and that the positronium breakup is roughly four times more probable 
than the hydrogen ionization. 

We have also obtained a simple formula for the total cross section for the positronium breakup 
which yields results in excellent agreement with the numerical computation. 

Finally we note that in fast collisions, 
where the momentum transfers are much smaller  
than the momentum of the incident projectile,  
the cross sections depend on the projectile velocity 
but are practically independent of the mass of the projectile. 
Therefore, the results of the present work 
can be directly applied also to the breakup/ionization 
by fast electrons (or positrons).

\section{ Acknowledgement }
 
We acknowledge the support from the National Key
Research and Development Program of China 
(Grant No. 2022YFA1602500), 
the CAS President’s International Fellowship Initiative, 
the Western Light - West Interdisciplinary Team 
Grant  No.  xbzgzdsys202406,  
and the National Laboratory of Heavy Ion Research Facility (HIRFL) in Lanzhou. 
A.B.V. is grateful for the hospitality of the
Institute of Modern Physics.


\begin{thebibliography}{99}


\bibitem{Ps-para-lifetime} S. G. Karshenboim, 
International Journal of Modern Physics {\bf A 19}, 3879–3896 (2003): arXiv:hep-ph/0310099 . 

\bibitem{Ps-ortho-lifetime-theor} G. S. Adkins, R. N. Fell, J. Sapirstein, 
Phys. Rev. Lett. {\bf 84}, 5086–5089 (2000): 
arXiv:hep-ph/0003028.

\bibitem{Ps-ortho-lifetime-exp} Y. Kataoka, S. Asai, T. Kobayashi, First Test of O(2) Correction of the Orthopositronium Decay Rate. Physics Letters {\bf B 671}, 219–223 (2009); arXiv:0809.1594. 


\bibitem{Ps-matter-1} C. Champion and C. Le Loirec, Phys. Med. Biol. {\bf 52}, 6605 (2007).  

\bibitem{Ps-matter-2} G. Garcia and M. C. Fuss (Editors), 
{\it Radiation Damage in Biomolecular Systems},  
(Springer Science, Business Media BV) (2012) .

\bibitem{Ps-matter-3} N. Guessoum, P. Jean and W. Gillard, 
Mon. Not. R. Astron. Soc. {\bf 402}, 1171, (2010). 

\bibitem{Ps-matter-4} G. Weidenspointner, C. R. Shrader, J. Kndlseder, P. Jean, 
V. Lonjou, N. Guessoum, R. Diehl, W. Gillard and M. J. Harris, Astron. Astrophys. {\bf 450}, 1013 (2006). 

\bibitem{Ps-matter-5} M. S. Briggs, et al., Geophys. Res. Lett. {\bf 38}, L02808 (2011).  

\bibitem{Ps-atom-review} G. Laricchia and H. R. J. Walters, 
Rivista del Nuovo Cimento {\bf 35}, N 6, 305 (2012).  

\bibitem{exper-with-Ps-1} D. B. Cassidy, P. Crivelli, T. H. Hisakado, L. Liszkay, V. E.
Meligne, P. Perez, H. W. K. Tom, and A. P. Mills, 
Phys. Rev. {\bf A 81}, 012715 (2010).

\bibitem{exper-with-Ps-2} K. Shu, Y. Tajima, R. Uozumi, N. Miyamoto, S. Shiraishi, T. Kobayashi, A. Ishida,
K. Yamada, R. W. Gladen, T. Namba, S. Asai, K. Wada, I. Mochizuki, T. Hyodo, K.
Ito, K. Michishio, B. E. O’Rourke, N. Oshima, K. Yoshioka1, 
Nature {\bf 633}, 793 (2024). 

\bibitem{exper-with-Ps-3} L. T. Gl\"oggler, 
Phys. Rev. Lett. {\bf 132}, 083402 (2024). 

\bibitem{Ps-atom-1} H. Ray, NIM 
{\bf B 192}, 191 (2002).  

\bibitem{Ps-atom-2} C. Starrett, Mary T. McAlinden, and H. R. J. Walters, 
Phys. Rev. {\bf A 72}, 012508 (2005).  

\bibitem{Ps-atom-3} S. Armitage, D.E. Leslie, J. Beale, G. Laricchia, 
NIM 
{\bf B 247}, 98 (2006). 

\bibitem{Ps-atom-4} H. Ray, {\bf EPL 73}, 21 (2006). 

\bibitem{Ps-atom-5} C. Starrett and H.R.J. Walters, 
Journal of Electron Spectroscopy and Related Phenomena {\bf 161}, 194 (2007).  

\bibitem{Ps-atom-6} S J Brawley and G Laricchia, J. Phys.: Conf. Ser. {\bf 199}, 012001 (2010).  
%

\bibitem{Ps-atom-8} D. Ghosh, S. Mukhopadhyay, and C. Sinha, 
Eur. Phys. J. D  67: 85 (2013)
%

\bibitem{c-Ps-1} Z. Kaliman, K. Pisk, and R. H. Pratt, Phys. Rev. {\bf A 83}, 053406 (2011). 

\bibitem{c-Ps-2} Z Kaliman and K Pisk, 
J. Phys. {\bf B 46} 235002 (2013). 


\bibitem{c-Ps-4} D. Dai and L. Fu, Phys. Rev. {\bf A 107}, 043118 (2023).

\bibitem{c-Ps-5} B. Najjari, C. Pedain, C. M\"uller, and A. B. Voitkiv, 
Phys. Rev. {\bf A 111}, 023109 (2025)


\bibitem{origin} Since we shall describe the center-of-mass motion of the target 
by plane waves, the exact location of the origin is not important (it should be just inside 
the interaction volume).   

\bibitem{ener-dist-e} 
R. Moshammer, W. Schmitt, J. Ullrich, H. Kollmus, A. Cassimi, 
R. D\"orner, O. Jagutzki, R. Mann, R. E. Olson et al, 
Phys. Rev. Lett. {\bf 79}, 3621 (1997). 

\bibitem{ener-dist-t} 
A. B. Voitkiv, B. Najjari, R. Moshammer 
and J. Ullrich, Phys. Rev. {\bf A 65}, 032707 (2002);    
A. B. Voitkiv and B. Najjari, 
J. Phys. {\bf B 37}, 4831 (2004).

\bibitem{E-M} J. Eichler and W. E. Meyerhof, {\it Relativistic Atomic Collisions },  
(San Diego: Academic Press, 1995). 

\bibitem{Eic} J. Eichler, {\it Lectures on Ion-Atom collisions} (Elsevier, New York 2005).   

\bibitem{Jack} J.D. Jackson, {\it Classical Electrodynamics}, 3rd ed., (Wiley,
New York, 1999).

\bibitem{abv2007} A. B. Voitkiv, J. Phys. {\bf B 40}, 2885 (2007).   
  
\bibitem{imp-par} Since the center-of-mass motion of the target 
is described by plane waves, in our treatment the vector $\bm b$ does not have 
the meaning of the projectile-target impact parameter which it has in 
the common form (see e.g. \cite{E-M}, \cite{Eic}) of the semi-classical approximation. 

\bibitem{b-d} J. D. Bjorker and S. D. Drell, 
{\it Relativistic Quantum Mechanics } (McGraw-Hill Book Company, New York, 1964). 

\bibitem{norm-vol} In obtaining (\ref{cr-1}) we have chosen $V_r = (2 \pi \hbar)^3 $. 

\bibitem{heavy-proton} The inspection of expressions (\ref{ampl-3}) and (\ref{ampl-4}) 
shows that the ionization of H via the projectile-proton interaction would require momentum transfers of the order of several hundreds of a.u. However, collisions with such momentum transfers are of negligible importance for the ionization of H.   

\bibitem{Bethe} H. Bethe, Ann. Phys. (Lpz.) {\bf 5}, 325 (1930). 

\bibitem{Inokuti} M. Inokuti, Rev. Mod. Phys. {\bf 43}, 297 (1971); 
Rev. Mod. Phys. {\bf 50}, 23 (1978). 

\bibitem{footnote} The relativistic effects at this energy remain very modest only provided one disregards the "trivial" difference between the relativistic and nonrelativistic dependences of the collision velocity on the impact energy. 




\end{thebibliography}
\end{document}